\pdfoutput=1
\documentclass[11pt]{article}
\usepackage[margin=1in]{geometry}
\usepackage{natbib}
\usepackage{needspace}
\usepackage{amsmath,amssymb}
\usepackage[utf8]{inputenc}
\usepackage[T1]{fontenc}
\usepackage{lmodern}
\usepackage[hyphens]{url}
\usepackage{hyperref}\hypersetup{hidelinks, pdftitle={When Sample-Splitting Choices Matter in TMLE and AIPW: Cross-Fitting, Targeting, and Repetition}}
\usepackage{graphicx}
\usepackage{booktabs}
\usepackage{longtable}
\usepackage{array}
\usepackage{caption}
\usepackage{listings}
\usepackage[most]{tcolorbox}
\newtcolorbox[auto counter]{biobox}[1][]{%
  breakable, colback=black!3, colframe=black!45, boxrule=0.4pt, arc=1pt,
  fonttitle=\bfseries, coltitle=black, left=4pt, right=4pt, top=3pt, bottom=3pt,
  title={Box~\thetcbcounter\ --- theoretical details}, #1}
\usepackage{enumitem}
\setlist[itemize]{topsep=2pt,partopsep=0pt,parsep=0pt,itemsep=1pt,leftmargin=1.4em}
\usepackage{setspace}
\lstdefinestyle{rstyle}{
  language=R, basicstyle=\ttfamily\footnotesize, breaklines=true,
  keywordstyle=\color{black}, commentstyle=\itshape\color{black!55},
  frame=single, rulecolor=\color{black!30}, showstringspaces=false,
  columns=fullflexible, xleftmargin=2pt, xrightmargin=2pt}
\providecommand{\tightlist}{\setlength{\itemsep}{0pt}\setlength{\parskip}{0pt}}
\providecommand{\passthrough}[1]{#1}
\newenvironment{keywords}{\par\medskip\noindent\textbf{Keywords:} }{\par\medskip}

\newcommand{\anon}{1}
\if1\anon\hypersetup{pdfauthor={M. Ehsan Karim and Rainie Fu}}\fi

\title{When Sample-Splitting Choices Matter in TMLE and AIPW:\\
Cross-Fitting, Targeting, and Repetition}

\if1\anon
\author{M. Ehsan Karim and Rainie Fu\\[4pt]
\normalsize
\begin{minipage}{0.9\textwidth}\centering
School of Population and Public Health, University of British Columbia,\\
2206 East Mall, Vancouver, BC V6T 1Z3, Canada\\[3pt]
Corresponding author: M. Ehsan Karim, \texttt{ehsan.karim@ubc.ca}\\[2pt]
ORCID (M. E. Karim): 0000-0002-0346-2871
\end{minipage}}
\else
\author{}
\fi
\date{}

\begin{document}
\maketitle

\begin{abstract}
Variants of targeted maximum likelihood estimation (TMLE) and augmented inverse-probability
weighting (AIPW) with machine-learning nuisance models differ in which observations train them,
how bias is corrected, and whether splits are repeated. In known-truth simulations of a
binary-outcome average treatment effect, eight headline estimators shared each setting's library,
inner folds and truncation bound; coverage differences were judged against a prespecified
two-percentage-point margin. With a low-complexity library and good overlap ($n=3000$), all but a
half-sample baseline covered equivalently (0.928--0.939). With one random forest on the primary
data-generating mechanism, estimators fitting the propensity in-sample undercovered severely
(0.12--0.48), whereas cross-fitting both nuisances raised coverage to 0.91--1.00, two overcovering.
Under near-positivity, every AIPW estimator covered better than every TMLE estimator with the
unweighted fluctuation (0.907--0.917 versus 0.656--0.806 at the primary setting, all
undercovering); looser truncation widened the gap. At the primary settings,
pooled-versus-fold-specific and single-versus-double cross-fit constructions (involving four
additional estimators) were usually equivalent; under extreme weights, pooled-targeting estimates were
1.7--1.8 times as dispersed as fold-specific ones. Repetition reduced partition dependence but made
several intervals more conservative. Two real-data analyses illustrate these sensitivities. Results
support cross-fitting both nuisances when empirical-process conditions cannot be justified, and
assessing truncation and AIPW-versus-TMLE sensitivity under poor overlap.
\end{abstract}


\setstretch{1.05}
\section{Introduction}\label{introduction}

\subsection{The problem an analyst
faces}\label{the-problem-an-analyst-faces}

An analyst who wants a doubly robust estimate of a treatment effect, by
targeted maximum likelihood estimation (TMLE) or augmented
inverse-probability weighting (AIPW), fits two nuisance models: a
propensity score and an outcome regression. With machine-learning fits,
several differently named estimators target the same effect through the
same estimating equation yet return different numbers. They differ in
four practical choices.

\begin{itemize}
\tightlist
\item
  \textbf{Which observations train the nuisance models, and which
  receive their predictions.} Some estimators fit and evaluate both
  models on the whole sample. Others hold data out (sample splitting or
  cross-fitting), so no observation's prediction comes from a model that
  saw it.
\item
  \textbf{Whether both nuisance models are held out, or only one.} Some
  estimators hold out the outcome model and fit the propensity score on
  everything.
\item
  \textbf{How bias is corrected.} AIPW puts the nuisance predictions
  straight into the estimating equation; TMLE first nudges the outcome
  predictions so the equation holds (the targeting step, pooled across
  folds or fold by fold).
\item
  \textbf{Whether the data are split once or many times}, and how
  repeated splits are combined.
\end{itemize}

An analyst who gets two answers from two packages cannot easily tell
whether the gap comes from these choices or from software defaults, such
as learner libraries, inner fold counts, and propensity-truncation
bounds. This paper holds those fixed and compares the choices.

The choices bear on whether the reported interval is trustworthy. With
low-complexity nuisance models and good overlap, the classical standard
error (SE) is usually adequate. With flexible learners it need not be:
the usual argument relies on a condition on the complexity of the fitted
nuisance functions that learners such as random forests and gradient
boosting are not guaranteed to satisfy
\citep{zheng2010asymptotic, balzer2021demystifying, naimi2021challenges}.
When that condition fails, bias and an understated variance can make the
nominal 95\% interval undercover. Holding data out is one way to avoid
relying on the condition (Section \ref{sec:estimators}).

\subsection{What this paper does}\label{what-this-paper-does}

Sample splitting reached applied researchers through the
targeted-learning \citep{zheng2010asymptotic}, econometrics
\citep{chernozhukov2018double} and epidemiology
\citep{zivich2021machine} literatures, each with its own vocabulary
(Supplement Section A). Three questions recur across them. Does
\textbf{cross-fitting} the nuisance models help, is cross-fitting the
outcome model alone enough, and does double cross-fitting improve on
single? Does the \textbf{bias correction} matter (AIPW versus TMLE;
pooled versus fold-specific targeting)? Does \textbf{repeating} the
split help? For each, we ask how large any difference is, with its Monte
Carlo uncertainty, and when it appears. None of the eight headline
estimators is new. What is new is a unified taxonomy and a controlled
comparison, under known truth, of specified TMLE and AIPW constructions
for the average treatment effect (ATE) on a binary outcome.

Within each setting, all estimators share the code, nuisance library,
inner cross-validation, truncation bound and, for TMLE, one unweighted
fluctuation (targeting step). Several comparisons still change more than
one design component (Section \ref{sec:estimators} says which), so not
every component's effect is separately identified.

\section{The estimators and their design space}\label{sec:estimators}

\subsection{Data, estimand, and the two nuisance
models}\label{data-estimand-and-the-two-nuisance-models}

We observe \(n\) independent copies of \(O=(W,A,Y)\): baseline
covariates \(W\), a binary treatment \(A\in\{0,1\}\), and a binary
outcome \(Y\in\{0,1\}\). The target is the ATE on the risk-difference
scale, \[
\psi \;=\; E\big[Y^{1}-Y^{0}\big]\;=\;E\big[Q_1(W)-Q_0(W)\big],\qquad Q_a(W)=E[Y\mid A=a,W],
\] where \(Y^a\) is the potential outcome under treatment \(a\). Two
nuisance functions appear throughout: the \textbf{outcome regression}
\(Q_a\) and the \textbf{propensity score} \(g(W)=P(A=1\mid W)\). Both
are estimated from data; everything that follows is about \emph{which}
data.

The estimand is identified under consistency, no unmeasured confounding
(\(Y^a \perp A \mid W\)), and positivity (\(0<g(W)<1\)); no estimator
choice can repair these assumptions.

Valid \emph{inference} needs more, and here the estimators part company.
The standard argument for regular root-\(n\) inference also assumes the
propensity bounded away from \(0\) and \(1\), consistent nuisance
estimators whose \(L_2\) errors have product \(o_P(n^{-1/2})\), and
control of an \emph{empirical-process} term, which for a nuisance fit
in-sample is usually obtained by limiting the fit's complexity
\citep{kennedy2022review}. Holding data out supplies that control
through independence instead; the others remain.

\subsection{One influence function, two
routes}\label{one-influence-function-two-routes}

The doubly robust estimators here build on the efficient influence
function (EIF), \[
D(O;g,Q,\psi)\;=\;\frac{A}{g(W)}\{Y-Q_1(W)\}-\frac{1-A}{1-g(W)}\{Y-Q_0(W)\}+Q_1(W)-Q_0(W)-\psi .
\] Given estimates of \(g\) and \(Q\), two routes solve
\(\mathbb{P}_n D=0\), where \(\mathbb{P}_n\) is the average over all
\(n\) observations. \textbf{AIPW}, the \emph{one-step} estimator, plugs
the initial estimates into \(D\) and solves for \(\psi\) directly.
\textbf{TMLE} instead updates \(\hat Q\) by a \emph{fluctuation} chosen
so the equation holds at the updated \(\hat Q^{*}\), then plugs
\(\hat Q^{*}\) in as a substitution estimator. Under the conditions
above, both are asymptotically linear with influence function \(D\),
hence asymptotically equivalent and efficient
\citep{kennedy2022review, hines2022}.

Supplement Section A gives the error decomposition containing the
empirical-process term and shows how splitting and cross-fitting act on
it. The plug-in (g-computation) comparators GCOMP and its cross-fit (CF)
form CF-GCOMP omit the augmentation terms of \(D\) and are reported
without an interval (Supplement Section H).

\subsection{Which data fit which model: a worked
fold}\label{which-data-fit-which-model-a-worked-fold}

Take \(n=3000\) and a five-fold split. Fold 1 holds observations
\(1\)--\(600\).

\begin{itemize}
\tightlist
\item
  For an observation in fold 1, a \emph{cross-fit} estimator fits
  \(\hat g\) and \(\hat Q\) on folds 2--5 (observations
  \(601\)--\(3000\)) and uses those fits to predict for observations
  \(1\)--\(600\). Rotating over all five folds gives every observation
  an out-of-fold prediction.
\item
  A \emph{no-split} estimator fits \(\hat g\) and \(\hat Q\) once on all
  \(3000\) and predicts for the same \(3000\). Observation \(i\)'s
  prediction comes from a model that saw it.
\item
  A \emph{three-way} scheme uses three folds of \(1000\). For each
  fold's observations (the estimation fold) it fits \(\hat g\) on one
  other fold and \(\hat Q\) on the third, so the two nuisance fits are
  also independent of each other.
\end{itemize}

\textbf{Two kinds of splitting, easily confused.} A Super Learner
chooses its ensemble weights by cross-validating \emph{inside} whatever
data it is given: \textbf{inner} cross-validation, a property of the
learner. The splitting above is \textbf{outer} cross-fitting: it decides
which data the learner is given and which observations receive its
predictions. Inner cross-validation does not make an estimator
cross-fit. We call all rotating outer splitting cross-fitting, including
the \(V\)-fold scheme of cross-validated TMLE (CV-TMLE), and vary only
the outer splitting.

\subsection{A taxonomy in five design
dimensions}\label{a-taxonomy-in-five-design-dimensions}

The dimensions (Table \ref{tab:axes}) are not fully separable:
fold-specific targeting needs held-out folds, and a two-way cross-fit is
a \(V\)-fold cross-fit with \(V=2\).

\begin{table}[htbp]\centering
\caption{The five design dimensions. An estimator is one choice on each dimension; not every combination is sensible, but most named estimators are one such combination.}
\label{tab:axes}
\small
\begin{tabular}{@{}>{\raggedright\arraybackslash}p{2.6cm}>{\raggedright\arraybackslash}p{4.4cm}>{\raggedright\arraybackslash}p{6.6cm}@{}}
\toprule
\textbf{Dimension} & \textbf{Choices} & \textbf{What it controls} \\
\midrule
1. Splitting & none; $V$-fold cross-fit; 2-way cross-fit; 3-way cross-fit & Whether $\hat g,\hat Q$ for observation $i$ are fit on data excluding $i$ \\
2. Nuisance location & both out-of-fold; outcome-only out-of-fold ($g$ in-sample) & Which nuisance is protected from own-observation overfitting \\
3. Bias correction & AIPW (one-step estimator); pooled TMLE fluctuation; fold-specific TMLE fluctuation & How $\mathbb{P}_nD=0$ is solved \\
4. Repetition & single split; $R$ repeated random splits & Reduces, without removing, dependence on any one random partition \\
5. Aggregation & none ($R=1$); median (with within+between variance) & How $R$ repeats are combined into one estimate and SE \\
\bottomrule
\end{tabular}
\end{table}

\textbf{Dimension 2 --- where the nuisances are fit.} Cross-fitting
\emph{only the outcome model} (CVQ-TMLE) leaves \(g\) in-sample.
\passthrough{\lstinline!tmle::tmle(cvQinit = TRUE)!} \citep{tmle}, the
\passthrough{\lstinline!tmle!} 2.1.1 default (with
\passthrough{\lstinline!automate=FALSE!}), does this approximately: its
per-fold outcome fits reuse the weights of a full-sample Super Learner.

\textbf{SS-TMLE} (split-sample TMLE) is the half-sample baseline. Both
nuisances are fit on one half and the estimator is evaluated on the
other, with \emph{no rotation}. It is an inefficient reference, not a
recommended estimator. The other seven estimators each use all \(n\)
observations for estimation (cross-fit forms included) and are called
\emph{full-sample}. Table \ref{tab:variants} lists the eight headline
estimators.

Four additional design cells (Supplement Section A) enter some class
means (mean coverage by design class) and Supplementary Figures S1--S2.
SCTMLE is a two-way single cross-fit TMLE \citep{zivich2024seed}: it
adds rotation to SS-TMLE's split and uses the fold-specific targeting of
double cross-fit TMLE (DCTMLE). The other three are \emph{mirror cells},
headline estimators with their targeting switched: CV-TMLE-fold,
CVQ-TMLE-fold and DC-TMLE-pool. CV-TMLE-fold corresponds in design to
the CVTMLE{[}all{]} of \citet{smith2024performance}, with a different
fluctuation submodel; the other two are unattested.

\begin{table}[!htbp]\centering
\caption{The eight headline estimators as choices in the design space. ``Bias correction'' is AIPW or a TMLE fluctuation, either pooled (one fluctuation $\epsilon=(\epsilon_0,\epsilon_1)$ across all evaluated observations) or fold-specific (a separate $\epsilon$ per fold, plug-ins averaged). Fold-averaging forms (fold-specific TMLE, DC-AIPW) use the mean within-fold variance and the others the stacked variance, so pooled-versus-fold-specific pairs and CF-AIPW versus DC-AIPW also switch the variance rule.}
\label{tab:variants}
\small
\resizebox{\ifdim\width>\linewidth\linewidth\else\width\fi}{!}{%
\begin{tabular}{@{}lllll@{}}
\toprule
\textbf{Label} & \textbf{Split} & \textbf{$g,Q$ fit} & \textbf{Bias correction} & \textbf{Key citation} \\
\midrule
AIPW         & none        & in-sample            & AIPW          & \citealp{robins1994} \\
TMLE         & none        & in-sample            & pooled        & \citealp{vanderlaan2006} \\
SS-TMLE      & 2-way, no rotation & one half        & pooled        & half-sample baseline \\
CVQ-TMLE     & $V$-fold    & $Q$ out, $g$ in      & pooled        & \citealp{smith2024performance}; \texttt{tmle} \\
CV-TMLE      & $V$-fold    & both out            & pooled        & \citealp{zheng2010asymptotic} \\
CF-AIPW      & $V$-fold    & both out            & AIPW          & \citealp{chernozhukov2018double} \\
DCTMLE       & 3-way double& both out (disjoint)  & fold-specific & \citealp{zivich2021machine} \\
DC-AIPW      & 3-way double& both out (disjoint)  & AIPW          & \citealp{newey2018cross, zivich2021machine} \\
\bottomrule
\end{tabular}%
}
\end{table}

\subsection{The algorithm and its variants}\label{sec:algorithm}

Vanilla TMLE takes six steps:

\begin{enumerate}
\def\labelenumi{\arabic{enumi}.}
\tightlist
\item
  \textbf{Fit the nuisances} \(\hat g\) and \(\hat Q\) on the whole
  sample.
\item
  \textbf{Truncate} \(\hat g\) away from \(0\) and \(1\); bound
  \(\hat Q\) so its logit is finite.
\item
  \textbf{Form the clever covariates} \(H_1=A/\hat g\) and
  \(H_0=(1-A)/(1-\hat g)\).
\item
  \textbf{Fit the fluctuation} \(\epsilon=(\epsilon_0,\epsilon_1)\) in
  \(\mathrm{logit}\,P(Y=1\mid A,W)=\mathrm{logit}\,\hat Q_A+\epsilon_0H_0+\epsilon_1H_1\)
  by unweighted logistic regression (the default of
  \passthrough{\lstinline!tmle!} 2.1.1 with
  \passthrough{\lstinline!automate = FALSE!}; versions 1.4.0--2.0.x
  defaulted to the weighted fluctuation, not studied here).
\item
  \textbf{Update and plug in:}
  \(\mathrm{logit}\,\hat Q_1^{*}=\mathrm{logit}\,\hat Q_1+\hat\epsilon_1/\hat g\),
  \(\mathrm{logit}\,\hat Q_0^{*}=\mathrm{logit}\,\hat Q_0+\hat\epsilon_0/(1-\hat g)\),
  and \(\hat\psi=\mathbb{P}_n(\hat Q_1^{*}-\hat Q_0^{*})\).
\item
  \textbf{Estimate uncertainty} from the influence function (below).
\end{enumerate}

Other estimators change these steps:

\begin{itemize}
\tightlist
\item
  \textbf{AIPW forms} (AIPW, CF-AIPW, DC-AIPW) replace steps 4--5 with
  \(\hat\psi=\mathbb{P}_n\{\hat Q_1-\hat Q_0+H_1(Y-\hat Q_1)-H_0(Y-\hat Q_0)\}\)
  \citep{robins1994}. TMLE's \(\hat\psi\), a mean of probability
  differences, stays in \([-1,1]\); the AIPW \(\hat\psi\) need not,
  though either Wald interval can cross these bounds \citep[Box
  10]{karim2026routes}.
\item
  \textbf{CV-TMLE} and \textbf{CF-AIPW} cross-fit both nuisances in step
  1, \textbf{CVQ-TMLE} only \(\hat Q\); \textbf{SS-TMLE} runs step 1 on
  one half and steps 2--6 on the other.
\item
  \textbf{DCTMLE} and \textbf{DC-AIPW} (double cross-fit) fit \(\hat g\)
  and \(\hat Q\) on disjoint folds, not on shared training folds as
  single cross-fit does, aiming at the remainder bias that shared
  training can add \citep[theory for AIPW-type
  estimators:][]{newey2018cross}. Both average the fold estimates, and
  DCTMLE fluctuates per fold. CF-AIPW versus DC-AIPW also changes fold
  count, training size and variance rule, so it is not an isolated
  single-versus-double contrast.
\end{itemize}

Supplement Section A gives R code for steps 2--5 and checks our
implementation against the \passthrough{\lstinline!tmle!} package
\citep{tmle}.

\subsection{Uncertainty, and what changes with
repetition}\label{uncertainty-and-what-changes-with-repetition}

All but the fold-averaging forms use the \emph{stacked} variance: the
empirical variance of the plugged-in influence function \(\hat D_i\)
divided by the number of observations averaged (\(n\), or \(n/2\) for
SS-TMLE). Fold-averaging forms (Table \ref{tab:variants}) divide the
\emph{mean within-fold} variance by \(n\) (Supplement Section A).

Under the \emph{median method} \citep{chernozhukov2018double}, the
estimate is the median of the \(R\) per-split estimates. The variance is
the median, over splits, of each split's within-split variance plus the
squared deviation of its estimate from the median estimate. That
\emph{same} variance must serve both the interval and any coverage
assessment; dropping the between-split term from either gives one
estimator two SEs.

\section{Simulation with known truth}\label{sec:sim}

We use data-generating mechanisms (DGMs) with known ATEs. Supplement
Section J summarizes the analysis plan (which prespecified the
comparison margins and decision rules), its reproducibility check and
the deviations from it.

\subsection{Design}\label{design}

\textbf{Data-generating mechanisms and regimes.} The primary DGM is the
statin confounding structure of \citet{zivich2021machine} (binary
outcome; true ATE \(\psi=-0.1081\)), adapted in R from the authors'
released code (Supplement Section B). Varying learner class and overlap,
two dimensions theory says should matter, gives three regimes:

\begin{itemize}
\tightlist
\item
  \textbf{S1 --- benign reference regime}: the statin DGM with a
  low-complexity Super Learner library (logistic regression, generalized
  additive model, sample mean).
\item
  \textbf{S2 --- a single random-forest nuisance fit}: the learner-class
  stressor, the statin DGM with one forest
  (\passthrough{\lstinline!ranger!}; its Super Learner weight is \(1\)).
\item
  \textbf{S3 --- near-positivity}: the overlap stressor, the statin DGM
  with the covariate part of the treatment model's logit multiplied by a
  sharpening factor (\(1\) recovers S1; larger values push propensities
  toward \(0\) and \(1\)), under the low-complexity library. At the
  primary factor \(2.0\), about \(11\%\) of the propensity fit's values
  fall outside \([0.025, 0.975]\) (about \(8.5\%\) at \(1.5\)), against
  about \(0.4\%\) in S1 (Supplement Section D and Table S6).
\end{itemize}

D3, a second DGM with a rough outcome surface (true ATE
\(\psi=+0.1007\)) fit with the S2 forest, checks whether the S2
\(g\)-splitting contrast is specific to the statin DGM (Supplement
Section C). Both true ATEs were computed by Monte Carlo and held fixed
across runs.

\textbf{Estimators and controls.} Every regime evaluates the eight
headline estimators of Table \ref{tab:variants}, plus the Supplement's
additional design cells and plug-in comparators. Within each setting
they share one Super Learner library \citep{vanderlaan2007super},
specified in advance following \citet{phillips2023practical}. Using one
library for all estimators and both nuisances is our comparability
control, not their advice, which is to tailor it to each prediction
task. The estimators also share a common inner cross-validation
(\(V=5\)) in every Super Learner fit (some packages default to 10), one
propensity-truncation bound (\([0.025,0.975]\) except in the S3 sweep)
and, for estimators with the same outer split, each dataset's partition,
drawn from a replicate-specific seed (Supplement Section J).

The outer splitting differs by design: \(V=5\) folds for CV-TMLE,
CVQ-TMLE and CF-AIPW, two for the single cross-fit SCTMLE and the
half-sample baseline, and three for double cross-fit. Double
cross-fitting fits each nuisance on \(n/3\) observations, so
cross-scheme contrasts also change training size.

\textbf{Scale, pairing and metrics.} Each regime's primary setting has
\(n=3000\) and 2500 simulated datasets; S1 and S2 are also run at
\(n=1000\) (Supplement Section E). S3 crosses two sharpening factors
(\(1.5\), \(2.0\)) with four truncation bounds (\(0.001\) to \(0.025\));
its primary setting is sharpening \(2.0\) with bound \(0.025\), and its
other seven cells use 1000 simulated datasets each. Common random
numbers give every estimator the \emph{same} datasets, so comparisons
are paired. The main comparisons use one split per dataset (\(R=1\)).

We report bias, 95\% coverage, the empirical SE (empSE, the standard
deviation of the estimates), the model-based SE (modSE, the root mean
square of the reported SEs), and
\(\mathrm{relSE}=100(\mathrm{modSE}/\mathrm{empSE}-1)\), with Monte
Carlo standard errors (MCSE) for coverage and bias
\citep{morris2019using}. With 2500 simulated datasets the coverage MCSE
is about \(0.4\) percentage points (pp) at nominal coverage and at most
\(1\) pp.

\textbf{A prespecified margin, not a significance test.} Because paired
comparisons over 2500 simulated datasets detect very small differences,
``materially different'' is judged against a prespecified margin, not a
test of zero difference. For paired coverage differences the margin is
\(\delta_{\mathrm{cov}}=0.02\). Two estimators are \emph{equivalent} if
the paired 90\% confidence interval for their coverage difference lies
inside \(\pm\delta_{\mathrm{cov}}\), the interval form of two one-sided
tests (TOST) at the 5\% level \citep{schuirmann1987}; they are
\emph{materially different} if it lies wholly outside
\(\pm\delta_{\mathrm{cov}}\), and otherwise \emph{inconclusive}. These
verdicts are relative to the chosen margin; they do not show that a
coverage difference inside it is unimportant in practice. Supplement
Table S9 gives the paired 90\% intervals for every headline pair at each
primary setting, so readers can apply other thresholds. An estimator
\emph{meets the prespecified coverage criterion} if the 90\% interval
for its own coverage lies inside \((0.93,0.97)\), \emph{undercovers} or
\emph{overcovers} if it lies wholly below \(0.93\) or above \(0.97\),
and is otherwise \emph{inconclusive}. The plan's \(0.005\) bias margin,
defined for paired differences, serves here only as a secondary
descriptive threshold for each estimator's absolute bias.

\subsection{S1 --- benign reference
regime}\label{s1-benign-reference-regime}

With a low-complexity nuisance library and good overlap, S1 is the
regime where the conditions for root-\(n\) inference are most plausible.
At \(n=3000\) the seven full-sample estimators of Table
\ref{tab:sim-headline} covered \(0.928\)--\(0.939\), and every pair was
equivalent within the \(0.02\) margin (Supplement Table S9; class means
in Supplement Section H).

\begin{table}[htbp]\centering
\caption{Two primary regimes at $n=3000$ (2500 simulated datasets): coverage of nominal 95\% intervals and relative SE error (\%). \textbf{Bold} marks the three in-sample estimators that undercovered severely under S2, and CF-AIPW, the only estimator there to meet the coverage criterion. Per-estimator bias and Monte Carlo SEs are in Supplement Table S7.}
\label{tab:sim-headline}
\small
\begin{tabular}{@{}llcrcr@{}}
\toprule
 & & \multicolumn{2}{c}{\textbf{S1 (benign)}} & \multicolumn{2}{c}{\textbf{S2 (forest)}} \\
\cmidrule(lr){3-4}\cmidrule(lr){5-6}
\textbf{Estimator} & \textbf{$g$ fit} & \textbf{cover} & \textbf{relSE} & \textbf{cover} & \textbf{relSE} \\
\midrule
TMLE      & in-sample  & 0.928 & $-5$  & \textbf{0.384} & $-62$ \\
AIPW      & in-sample  & 0.932 & $-4$  & \textbf{0.483} & $-52$ \\
CVQ-TMLE  & in-sample ($Q$-only) & 0.929 & $-5$ & \textbf{0.124} & $-61$ \\
SS-TMLE   & split-half & 0.918 & $-7$  & 0.999 & $+128$ \\
CV-TMLE   & cross-fit  & 0.933 & $-3$  & 0.999 & $+125$ \\
CF-AIPW   & cross-fit  & 0.936 & $-2$  & \textbf{0.938} & $+22$ \\
DCTMLE    & cross-fit  & 0.935 & $-2$  & 0.993 & $+96$ \\
DC-AIPW   & cross-fit  & 0.939 & $-2$  & 0.911 & $+9$ \\
\bottomrule
\end{tabular}
\end{table}

Every estimator covered below the nominal \(0.95\), the seven
full-sample ones by \(1\)--\(2\) pp.~No splitting scheme or correction
route removed the shortfall, and its source was not identified. Of the
eight, only DC-AIPW met the coverage criterion with 2500 simulated
datasets; SS-TMLE fell below it, and the rest were inconclusive. SS-TMLE
covered lowest (\(0.918\)), although its reported SE grew roughly in
proportion to its larger empirical SE (\(0.027\) against
\(0.018\)--\(0.019\)), the expected cost of evaluating on half the data.
SS-TMLE was equivalent to vanilla TMLE but, with 2500 simulated
datasets, inconclusive against the other six.

\subsection{S2 --- a single random-forest nuisance
fit}\label{s2-a-single-random-forest-nuisance-fit}

Replacing the low-complexity library with a single random forest
separates the estimators by where the propensity score is fit (Table
\ref{tab:sim-headline}; Supplementary Figure S5). The three that fit it
\textbf{in-sample} undercovered severely (\(0.12\)--\(0.48\)), the
outcome-only CVQ-TMLE worst; those that \textbf{cross-fit both
nuisances} did not (class means in Supplement Section H). The driving
contrast holds the \(Q\)-fitting procedure fixed: CVQ-TMLE and CV-TMLE
are both pooled \(V\)-fold estimators with \(Q\) cross-fit, and differ
in procedure \emph{only} in where the propensity is fit (learner
randomness can still make their fitted \(Q\) differ). They covered
\(0.124\) and \(0.999\).

Cross-fitting raised coverage to \(0.91\)--\(1.00\) but did not by
itself deliver calibrated intervals. Every cross-fit TMLE form, the
additional cells included, overcovered (\(0.99\)--\(1.00\), with
model-based SE inflated by \(86\)--\(125\%\); Supplement Section H), as
did SS-TMLE (\(0.999\)). Only CF-AIPW met the coverage criterion
(\(0.938\), at its lower edge), and DC-AIPW still undercovered slightly
(\(0.911\)). Mechanism diagnostics (Supplement Section G) show a
pattern, not an identified mechanism.

This cell deliberately uses a single random forest at Super Learner
weight \(1\), to make in-sample optimism large; mixed libraries, in
which Super Learner may down-weight the forest, were not tested. Severe
undercoverage is thus established for this learner when the propensity
is fit in-sample, not for machine-learning libraries generally.

\textbf{Independent check (D3).} On the second, rough-surface DGM
(Supplement Section C) the same pattern recurred: severe in-sample-\(g\)
undercoverage (\(0.02\)--\(0.41\)) and recovery under cross-fitting
(\(0.947\)--\(0.956\), additional cells included). Unlike S2, every
cross-fit form met the coverage criterion.

\subsection{S3 --- near-positivity}\label{s3-near-positivity}

When overlap rather than the learner is degraded and TMLE uses the
unweighted fluctuation of step 4, the correction route (AIPW versus
TMLE) mattered more than nuisance splitting. At the primary setting
every AIPW form (\(0.907\)--\(0.917\); mean \(0.912\) over three)
covered better than every TMLE form (\(0.656\)--\(0.806\); mean
\(0.779\) over nine, additional cells included; Supplement Tables S7 and
S2), though each undercovered. The gap widened as the bound was
loosened: at the same sharpening and bound \(0.001\) the AIPW and TMLE
means were \(0.926\) and \(0.440\) (Supplement Table S2). For the
cross-fit AIPW forms this was an advantage in coverage, not precision
(Supplement Section H).

Three matched pairs each share one nuisance fit: AIPW/TMLE,
CF-AIPW/CV-TMLE and DC-AIPW/DCTMLE. Within each pair, the gap therefore
arises in the bias-correction step rather than the nuisance fits. Across
the eight S3 settings, the AIPW member covered materially better in 23
of these 24 pair-by-setting comparisons, and the remaining one was
inconclusive; the comparisons are related, not independent. At the
primary setting the reported SEs within each pair differed by under
\(3\%\), but the TMLE estimates were \(1.32\)--\(1.38\) times as
dispersed (Supplement Section G, in the text introducing Table S7, notes
one untested explanation).

Among full-sample estimators, nuisance splitting was secondary. With the
correction route held fixed, the \(g\) contrast (TMLE pair: CVQ-TMLE
versus CV-TMLE, both cross-fitting \(Q\), \(0.786\) versus \(0.789\);
AIPW pair: AIPW versus CF-AIPW, which changes \(Q\) too) was equivalent
within the \(0.02\) margin at the primary setting for both routes.
Across the sweep it was material only at sharpening \(1.5\) with the
loosest bounds, and always smaller than the same cell's AIPW-versus-TMLE
gap (Supplement Section D, Table S3).

\subsection{Finer choices and
repetition}\label{finer-choices-and-repetition}

At the primary settings the five pooled-versus-fold-specific and
single-versus-double pairs (four involve an additional design cell;
Supplement Section H) were equivalent within the margin in 16 of 20
comparisons. The single-versus-double pairs also change fold count and
training size. On S2, pooled CVQ-TMLE (\(0.124\)) was materially worse
than CVQ-TMLE-fold (\(0.272\)), both failing severely. In the most
extreme S3 cell (sharpening \(2.0\), bound \(0.001\)) the pooled forms
had \(1.67\)--\(1.83\) times the fold-specific forms' empirical SE and
\(32\)--\(34\) pp lower coverage. Both forms undercovered there, and
each pair also switches the variance rule, whose share of the coverage
gap was not isolated.

For repetition, each dataset's \(R\) splits were combined by the median
method before coverage was computed across datasets. Repetition reduced
single-partition noise in the point estimate but did not calibrate the
intervals. On S1 (500 simulated datasets, \(R\le100\)) only SS-TMLE's
empirical SE changed appreciably, settling by \(R\approx5\)--\(10\) (a
dataset-averaged plateau, not a recommended \(R\); Supplement Section
F). On S2 (250 simulated datasets; coverage MCSE up to about \(2\) pp)
the empirical SE fell with \(R\), but the reported SE did not: its
within-split part stayed flat. The cross-fit AIPW forms, whose
between-split term added \(5\)--\(13\%\), moved from below nominal to at
or above it (CF-AIPW \(0.908\to0.964\), DC-AIPW \(0.912\to0.992\) from
\(R=1\) to \(R=25\); Supplement Section F).

\section{Real-data illustrations}\label{sec:realdata}

\subsection{Good overlap: right heart
catheterization}\label{good-overlap-right-heart-catheterization}

Two analyses ask whether the choices that mattered in simulation also
move real-data estimates and standard errors; the truth is unknown. The
first uses the right-heart-catheterization (RHC) cohort
\citep{connors1996}: \(n=5735\) critically ill adults, of whom \(2184\)
received RHC within 24 hours of admission to intensive care. The target
is the ATE of RHC on 30-day mortality (risk difference), adjusting for
the standard set of 49 covariates. Overlap is good: only \(1.9\%\) of
patients have an estimated propensity outside the shared
\([0.025, 0.975]\) truncation bound (Supplement Section L). A
\emph{basic} Super Learner library (logistic regression and the sample
mean) is compared with an \emph{extended} one that adds a random forest.
Split estimators are median-aggregated over \(R=10\) repetitions under
the basic library and use a single split under the extended one.

Under the basic library (Supplementary Figure S1), every full-sample
estimator places the ATE between \(0.059\) and \(0.064\), with almost
fully overlapping 95\% intervals. The half-sample SS-TMLE reads
\(0.067\), with the widest interval. Every estimate is positive (higher
30-day mortality with RHC), as \citet{connors1996} reported. As in S1
(also a low-complexity library), the full-sample estimators agree.
Agreement is not correctness: another analysis of this cohort and
estimand \citep{karim2026routes} reports smaller estimates (Supplement
Section L).

Adding a random forest to the ensemble moves the no-split estimates and
standard errors (Supplement Table S10). Estimates then spread from
\(0.035\) (the plug-in GCOMP) to \(0.099\) (CVQ-TMLE), while the seven
fully cross-fit estimators, additional design cells included, read
\(0.050\) to \(0.057\). The standard errors of the no-split estimators
(vanilla TMLE, AIPW), fit once under each library, fall from about
\(0.013\) to \(0.005\) (Supplementary Figure S2). CVQ-TMLE's SE
(propensity in-sample) falls to \(\approx0.008\), and the fully
cross-fit ones only slightly, to about \(0.012\). Within the
extended-library run (one fit or split per estimator), standard errors
thus track where the nuisances are fit, not the bias correction. For the
split estimators the two runs also differ in \(R\), SE rule and
partition, so the library's effect on them is not isolated. None of this
measures coverage.

\subsection{Poor overlap: the LaLonde
benchmark}\label{poor-overlap-the-lalonde-benchmark}

The second illustration examines poor overlap against a randomized
benchmark, the National Supported Work (NSW) job-training experiment. In
the Dehejia--Wahba subsample of LaLonde's male participants {[}\(185\)
treated, \(260\) controls; \citet{dehejia1999}{]}, the experimental
effect on employment in 1978 is a risk difference of \(+0.111\).
Following \citet{lalonde1986}, we replace the experimental controls with
a comparison group from the Current Population Survey (CPS) or the Panel
Study of Income Dynamics (PSID), and estimate the ATE on employment over
the pooled sample, adjusting for eight baseline covariates that include
1974 and 1975 earnings.

Two caveats apply to any comparison with \(+0.111\). First, the
benchmark averages over NSW enrollees (Dehejia and Wahba target the
effect on the treated), whereas our estimand averages over the pooled
sample, mostly comparison units (\(99\%\) with CPS, \(93\%\) with PSID).
Second, the comparison assumes no unmeasured confounding given these
covariates, an assumption to which Dehejia and Wahba found their own
estimates sensitive. Distance from the benchmark is therefore not bias
in the usual sense, nor attributable to positivity alone (Supplement
Section L).

The estimated propensity scores (Supplementary Figure S3) show a
practical positivity violation before any outcome is modeled: about
\(90\%\) of CPS comparison units (and \(78\%\) of PSID's) have an
estimated propensity below \(0.01\). We compare vanilla TMLE, AIPW,
CV-TMLE and DCTMLE (median over \(R=20\) splits), all with the basic
library and the \([0.025, 0.975]\) bound, which truncates the no-split
propensity for \(95\%\) of the pooled sample with CPS and \(80\%\) with
PSID.

With CPS, whose overlap is worse, all four estimates lie below the
benchmark and none of their intervals reaches it (Supplement Table S11;
PSID results are in Supplement Section L). The three TMLE forms read
\(+0.069\) to \(+0.077\) and AIPW \(+0.011\). Vanilla TMLE and AIPW
share their nuisance fits, so this gap of about \(0.06\), under the
\emph{same} bound and library, lies entirely in the bias correction, the
dimension that mattered under near-positivity in S3. Without this
target's truth, the gap shows only that the estimator choice changed the
answer, not which estimate is closer.

\begin{figure}[!tb]\centering
\includegraphics[width=5in,alt={TMLE estimate against propensity-truncation bound; tighter truncation moves the estimate further from the randomized benchmark and narrows the interval so it no longer reaches it.}]{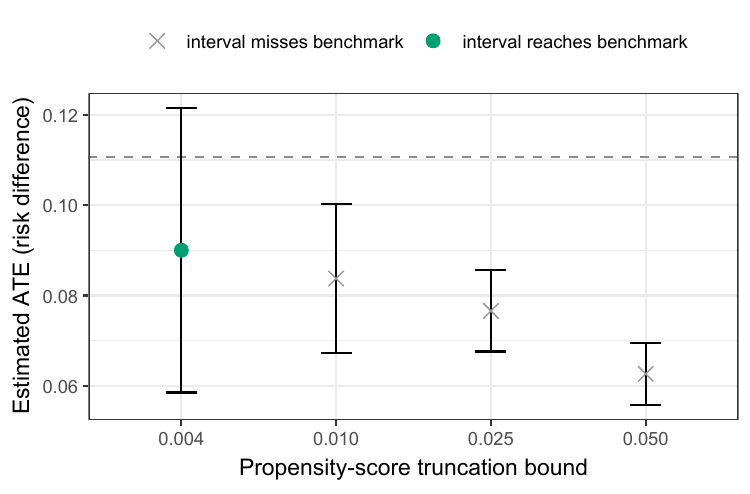}
\caption{Vanilla TMLE estimate of the LaLonde employment effect (CPS comparison) with 95\% intervals, by propensity-truncation bound; dashed line, the randomized benchmark, which only the loosest bound's interval reaches. The benchmark is the effect among NSW enrollees, not this estimand's truth.\label{fig:lalonde-trunc}}
\end{figure}

The truncation bound, often left at a default, also moves the answer
(Figure \ref{fig:lalonde-trunc}: vanilla TMLE, CPS comparison).
Tightening the bound from \(0.004\) (the default lower bound of
\texttt{tmle} 2.1.1 with \texttt{automate = FALSE} at this \(n\)) to
\(0.05\) moves the point estimate from \(+0.090\) to \(+0.063\), away
from the benchmark, and shrinks the standard error more than fourfold.
The bound moved this estimate (\(0.027\)) more than the choice among the
three TMLE forms did (\(0.008\)). Under poor overlap, a narrow interval
is no reassurance; its width does not reflect sensitivity to the
truncation bound.

\section{Practical guidance and reporting}\label{sec:reporting}

\subsection{An operational rule}\label{an-operational-rule}

\textbf{Diagnose overlap first}: on the LaLonde data the estimate moved
with both the estimator and the truncation bound. Table \ref{tab:rule}
asks four questions.

\begin{table}[!htb]\centering
\caption{Four questions to ask before choosing a sample-splitting TMLE or AIPW estimator, with the evidence behind each answer.}
\label{tab:rule}
\resizebox{\ifdim\width>\linewidth\linewidth\else\width\fi}{!}{%
\begin{tabular}{@{}>{\raggedright\arraybackslash}p{2.2cm}>{\raggedright\arraybackslash}p{4.6cm}>{\raggedright\arraybackslash}p{8.6cm}@{}}
\toprule
\textbf{Ask} & \textbf{Diagnose} & \textbf{Then} \\
\midrule
Overlap? & From a low-complexity propensity fit, inspect the histogram, range and share outside $[0.025,0.975]$; near-positivity? & \textbf{Assess sensitivity} to propensity truncation and to the correction route (AIPW versus TMLE). Report estimates, standard errors and the fraction truncated. In S3, AIPW covered better than unweighted-fluctuation TMLE but still undercovered at the primary setting. Unlike TMLE's, an AIPW estimate can leave $[-1,1]$; check for it. \\
\addlinespace
Learner class? & Can empirical-process conditions be justified for the chosen learners? Random forests and boosting, for example, are not guaranteed to satisfy them. & \textbf{Cross-fit both nuisances} if not (the simulation tested only a single random forest). In both single-forest regimes this raised coverage to $0.91$--$1.00$ without guaranteeing calibration: on S2 the cross-fit TMLE forms overcovered and DC-AIPW undercovered ($0.911$). \\
\addlinespace
Software default? & \texttt{tmle::tmle()} with \texttt{cvQinit=TRUE} (the \texttt{tmle} 2.1.1 default with \texttt{automate=FALSE}) cross-fits $Q$ but leaves the propensity in-sample (a CVQ-TMLE design). & Do \emph{not} rely on it when the Learner row calls for cross-fitting: despite cross-fitting $Q$, our CVQ-TMLE had the worst coverage in both forest regimes ($0.02$--$0.12$ at $n=3000$). At the S3 primary setting it was not uniquely bad; the half-sample SS-TMLE was worst (Supplement Table S7). \\
\addlinespace
Which cross-fit form? & After cross-fitting both nuisances, is any one form always best? & \textbf{No universal ranking.} Every cross-fit form met the coverage criterion on D3, but on S2 only CF-AIPW did, with a point bias of $-0.029$ (beyond the $0.005$ threshold) and an empirical SE about twice CV-TMLE's ($0.040$ against $0.022$). \\
\bottomrule
\end{tabular}%
}
\end{table}

\subsection{What to report}\label{what-to-report}

Table \ref{tab:checklist} lists the minimum a reader needs to
reconstruct the design choices. Its generic items, splitting scheme
included, adapt the checklist of \citet{karim2026routes}; we add the
other splitting-design items. Supplement Section F discusses the seed
row and how many repetitions to use.

\begin{table}[htbp]\centering
\caption{Reporting checklist for a sample-splitting TMLE or AIPW analysis.}
\label{tab:checklist}
\small
\begin{tabular}{@{}>{\raggedright\arraybackslash}p{3.1cm}p{12.6cm}@{}}
\toprule
\textbf{Item} & \textbf{What to state} \\
\midrule
Estimand & ATE (risk difference / ratio), target population, scale \\
Identification & consistency, exchangeability (the adjustment set), positivity (and how checked) \\
Super Learner & the full library; inner cross-validation fold count; screening \\
Splitting & none / half-sample / $V$-fold, 2-way or 3-way cross-fit; number of folds; both $g$ and $Q$ cross-fit, or $Q$ only \\
Bias correction & AIPW, pooled or fold-specific TMLE fluctuation; the fluctuation submodel (covariates, weights) \\
Truncation & propensity bounds; the \% of observations truncated; under poor overlap, how estimates and SEs change across bounds and between AIPW and TMLE \\
Repetition & number of splits $R$; aggregation rule; within- and between-split variance \\
Seed & how the seeds or partitions were chosen, and whether results are sensitive to them \\
Software & package versions; non-default arguments (e.g., \texttt{cvQinit}, \texttt{target.gwt}, \texttt{automate}) \\
Uncertainty & analytic SE and its variance rule (stacked or within-fold); for simulations, the Monte Carlo SE of each metric \citep{morris2019using} \\
\bottomrule
\end{tabular}
\end{table}

\section{Discussion}\label{discussion}

Which design dimension mattered most depended on the regime. In the
benign regime the full-sample estimators were equivalent within the
\(0.02\) margin; with a random forest as the sole candidate learner,
cross-fitting both nuisances mattered; under near-positivity, the
correction route (AIPW versus TMLE) did (Supplement Section N). Every
repeated analysis used the median method, so aggregation rules were not
compared. Supplement Section I relates these results to the simulation
study of \citet{ellul2025crossfit}, which slightly favored TMLE under a
different design; neither study establishes a universal preference under
poor overlap.

\textbf{Scope.} We studied only the binary-outcome ATE, the five
dimensions of Table \ref{tab:axes}, fixed truncation bounds and one
fluctuation (the unweighted one of step 4, Section \ref{sec:algorithm}),
leaving aside collaborative TMLE, the highly adaptive lasso
\citep{vanderlaan2023hal} (which can control the empirical-process term
without splitting) and undersmoothing, higher-order influence-function
estimators, and longitudinal, survival and survey-weighted extensions.

\textbf{Limitations.} The known-truth evidence (Section \ref{sec:sim})
comes from two data-generating mechanisms, mainly at \(n=3000\): the
statin-style one, with near-positivity simulated only under the
low-complexity library, and an independent rough-surface one. What we
claim is qualitative: which contrasts between the estimators studied
were material in which regime, not a universal ranking or exact coverage
outside these regimes. The two real datasets (Section
\ref{sec:realdata}) \emph{illustrate} sensitivity only; they cannot
validate coverage.

Under the random-forest fits, cross-fitting the propensity as well as
the outcome model removed the severe undercoverage (CV-TMLE versus
CVQ-TMLE, matched in outcome-fitting procedure but not necessarily in
fitted predictions), but why in-sample propensity fits degrade the
intervals was not isolated. Under extreme weights, the variance rule's
share of the pooled-versus-fold-specific coverage gap was not isolated
either. We do not prove that the three mirror cells (headline estimators
with their targeting switched) are consistent; their variances follow
attested conventions. Finally, doubly robust estimates with flexible
learners retain a seed and partition dependence that repetition
mitigates but does not remove.

\textbf{Conclusion.} Cross-fit both nuisances when empirical-process
conditions cannot be justified for the chosen learners. Diagnose overlap
and assess sensitivity to truncation and to the correction route (AIPW
versus TMLE). In our near-positivity simulation (low-complexity
library), AIPW estimators had higher coverage than
unweighted-fluctuation TMLE estimators; the preferred estimator still
depends on the setting and performance criterion.

\newcommand{\closinghead}[1]{\par\medskip\noindent\textbf{#1.}\enspace}

\closinghead{Supplementary material}
The Supplementary Materials (PDF) give the estimator reference, the data-generating mechanisms as
implemented, further simulation results and a regime summary (Tables S1--S7 and S12, Figures S4--S5), the
analysis plan (Supplement Section J) and software environment (Table S8), paired coverage comparisons (Table S9), and the real-data estimates and figures
(Tables S10--S11, Figures S1--S3).
\if1\anon
The code and saved results are in the companion repository described below.
\else
An anonymized copy of the companion repository (ZIP) holds the code and saved results described below.
\fi

\closinghead{Data and code availability}
The right heart catheterization data are those of \citet{connors1996}. Data obtained from
\url{http://hbiostat.org/data} courtesy of the Vanderbilt University Department of Biostatistics.
The NSW, CPS and PSID files \citep{lalonde1986,dehejia1999,dehejia2002} were obtained from
\url{https://users.nber.org/~rdehejia/nswdata2.html}, which distributes them for attributable
non-commercial use. The companion repository holds the R code, including an R adaptation of the
data-generating mechanism of \citet{zivich2021machine} based on their released code \citep{zivich2026code},
which its release record licenses under CC~BY~4.0; the saved simulation results; and summary results of the
applied analyses, but no benchmark data. From the saved results one script regenerates
Table~\ref{tab:sim-headline}, Figure~\ref{fig:lalonde-trunc}, Supplementary Figures S1, S2 and S5 and Supplementary Tables S1--S7 and S9--S11 without model fitting (Table S11's sample and overlap
summaries are carried as reported); Supplementary Figure S3 and any refit need the data, which the code
downloads.
\if1\anon
The repository is \url{https://github.com/ehsanx/TMLE-variants-reproduction}; an archived DOI will
be added on release.
\else
An anonymized copy of the repository is provided as supplementary material.
\fi

\if1\anon
\closinghead{Acknowledgments}
This research was supported in part through computational resources and services provided by
Advanced Research Computing at the University of British Columbia.
\fi

\closinghead{Declaration of generative AI use}
\if1\anon
During this work, the authors used AI-based tools (large language models) to assist with the analysis and
simulation code, and text editing; the authors verified all outputs and take full responsibility for the
content.
\else
During this work, the authors used AI-based tools (large language models) to assist with the analysis and
simulation code, and text editing; the authors verified all outputs and take full responsibility for the
content.
\fi

\closinghead{Funding}
This research received no specific grant from any funding agency in the public,
commercial, or not-for-profit sectors.

\closinghead{Disclosure statement}
\if1\anon The authors report there are no competing interests to declare.\else The authors report there are no competing interests to declare.\fi

\bibliographystyle{agsm}
\bibliography{ref}
\end{document}